\documentclass[sigconf]{acmart}

\acmConference[ICSE 2022]{The 44th International Conference on Software Engineering}{May 21–29, 2022}{Pittsburgh, PA, USA}

\AtBeginDocument{%
\providecommand\BibTeX{{%
\normalfont B\kern-0.5em{\scshape i\kern-0.25em b}\kern-0.8em\TeX}}}

\copyrightyear{2022}
\acmYear{2022}
\setcopyright{acmcopyright}\acmConference[ICSE '22 Companion]{44th International
Conference on Software Engineering Companion}{May 21--29, 2022}{Pittsburgh, PA,
USA}
\acmBooktitle{44th International Conference on Software Engineering Companion
(ICSE '22 Companion), May 21--29, 2022, Pittsburgh, PA, USA}
\acmPrice{15.00}
\acmDOI{10.1145/3510454.3516839}
\acmISBN{978-1-4503-9223-5/22/05}

\usepackage{stfloats}
\usepackage{paralist}
\usepackage{braket}
\usepackage{nicefrac}
\usepackage{xcolor}
\usepackage{multirow}
\usepackage{verbatim}
\usepackage{graphicx}
\usepackage{xspace}
\usepackage{booktabs}

\theoremstyle{definition}
\newtheorem{definition}{\protect\definitionname}
\providecommand{\definitionname}{Definition}

\newcommand{\quantumBitsSet}{\ensuremath{Q}\xspace}
\newcommand{\quantumProg}{\ensuremath{\textsc{QP}}\xspace}
\newcommand{\inputsSet}{\ensuremath{I}\xspace}
\newcommand{\outputsSet}{\ensuremath{O}\xspace}
\newcommand{\domain}[1]{\ensuremath{D_{#1}}\xspace}
\newcommand{\inputDom}{\ensuremath{\domain{I}}\xspace}
\newcommand{\outputDom}{\ensuremath{\domain{O}}\xspace}

\newcommand{\progSpec}{\ensuremath{\mathtt{PS}}\xspace}
\newcommand{\progSpecIO}[2]{\ensuremath{\progSpec(#1, #2)}\xspace}

\newcommand{\testInputValue}{\ensuremath{i}\xspace}
\newcommand{\fail}{\ensuremath{\mathtt{fail}}\xspace}

\newcommand{\res}{\ensuremath{\mathit{res}}\xspace}

\newcommand{\percInputs}{\ensuremath{\beta}\xspace}

\newcommand{\uof}{\textit{uof}\xspace}
\newcommand{\wodf}{\textit{wodf}\xspace}
\newcommand{\numRepIndex}{\ensuremath{n}\xspace}
\newcommand{\numTests}{\ensuremath{M}\xspace}

\newcommand{\qusbt}{\texttt{QuSBT}\xspace}

\newcommand{\addSName}{\ensuremath{\mathtt{AS}}\xspace}
\newcommand{\bvName}{\ensuremath{\mathtt{BV}}\xspace}
\newcommand{\ceName}{\ensuremath{\mathtt{CE}}\xspace}
\newcommand{\qramName}{\ensuremath{\mathtt{QR}}\xspace}
\newcommand{\iqftName}{\ensuremath{\mathtt{IQ}}\xspace}

\newcommand{\firstFaultyQP}{FQP$_1$\xspace}
\newcommand{\secondFaultyQP}{FQP$_2$\xspace}

\newcommand{\percFailingTests}{\%ft\xspace}
\newcommand{\numTestCases}{M\xspace}
\newcommand{\simulationTime}{st\xspace}
\newcommand{\executionTime}{et\xspace}

\copyrightyear{2022}
\acmYear{2022}
\setcopyright{acmcopyright}\acmConference[ICSE '22 Companion]{44th International
Conference on Software Engineering Companion}{May 21--29, 2022}{Pittsburgh, PA,
USA}
\acmBooktitle{44th International Conference on Software Engineering Companion
(ICSE '22 Companion), May 21--29, 2022, Pittsburgh, PA, USA}
\acmPrice{15.00}
\acmDOI{10.1145/3510454.3516839}
\acmISBN{978-1-4503-9223-5/22/05}

\begin{document}

\title{QuSBT: Search-Based Testing of Quantum Programs}


\author{Xinyi Wang}
\email{wangxinyi125@nuaa.edu.cn}
\orcid{0000-0001-5621-6140}
\affiliation{%
\institution{Nanjing University of Aeronautics and Astronautics}
\city{Nanjing}
\country{China}
}

\author{Paolo Arcaini}
\email{arcaini@nii.ac.jp}
\orcid{0000-0002-6253-4062}
\affiliation{%
\institution{National Institute of Informatics}
\city{Tokyo}
\country{Japan}
}

\author{Tao Yue}
\email{taoyue@ieee.org}
\orcid{0000-0003-3262-5577}
\affiliation{%
\institution{Nanjing University of Aeronautics and Astronautics}
\institution{Simula Research Laboratory, Norway}
\city{Nanjing}
\country{China}
}

\author{Shaukat Ali}
\email{shaukat@simula.no}
\orcid{0000-0002-9979-3519}
\affiliation{%
\institution{Simula Research Laboratory}
\city{Oslo}
\country{Norway}
}


\begin{abstract}
Generating a test suite for a quantum program such that it has the maximum number of failing tests is an optimization problem. For such optimization, search-based testing has shown promising results in the context of classical programs. To this end, we present a test generation tool for quantum programs based on a genetic algorithm, called \qusbt (Search-based Testing of Quantum Programs). \qusbt automates the testing of quantum programs, with the aim of finding a test suite having the maximum number of failing test cases. \qusbt utilizes IBM's Qiskit as the simulation framework for quantum programs. We present the tool architecture in addition to the implemented methodology (i.e., the encoding of the search individual, the definition of the fitness function expressing the search problem, and the test assessment w.r.t. two types of failures). Finally, we report results of the experiments in which we tested a set of faulty quantum programs with \qusbt to assess its effectiveness.\\
\noindent Repository (code and experimental results): \url{https://github.com/Simula-COMPLEX/qusbt-tool}\\
Video: \url{https://youtu.be/3apRCtluAn4}
\end{abstract}

\ccsdesc[500]{Theory of computation~Quantum computation theory}
\ccsdesc[500]{Software and its engineering~Software testing and debugging}
\ccsdesc[500]{Software and its engineering~Search-based software engineering}

\keywords{Quantum Programs, Search-Based Testing, Genetic Algorithms}


\maketitle

\section{Introduction}\label{sec:introduction}

Quantum Computing (QC) promises to bring several advantages by solving intrinsically complex problems. As for classical programs, testing of quantum programs is essential to guarantee their correctness. However, the probabilistic nature and the specific features of QC, such as superposition and entanglement, make testing quantum programs very challenging.

By realizing the importance of quantum software testing~\cite{MiranskyyICSE19,ERCIM2022,zhao2020quantum}, several research works have been proposed. Notable examples include property-based testing~\cite{honarvar2020property}, fuzz testing~\cite{quantfuzzposterICST2021}, mutation analysis~\cite{Muskit}, search-based testing~\cite{genTestsQPSSBSE2021}, combinatorial testing~\cite{CTQuantumQRS2021}, input/output coverage~\cite{ourICST2021} along with the \texttt{Quito} tool~\cite{quitoASE21tool}, and specialized projection-based assertions~\cite{Li2020} for quantum programs.

However, in order for these techniques to be adopted in the development practice, the automation and tool support of the testing process are essential. Towards these goals, we present the \qusbt (Search-based Testing of Quantum Programs) testing tool for quantum programs, which employs a Genetic Algorithm (GA) to automatically generate a test suite with the maximum possible number of failing test cases. Note that the approach used by \qusbt to generate such test suites was published in our previous work~\cite{genTestsQPSSBSE2021}, but no tool usable by users was provided there. Therefore, in this paper, we present the \qusbt tool, which is implemented for quantum programs coded using IBM's Qiskit~\cite{QiskitWille2019} framework in Python. \qusbt also relies on the jMetalPy~\cite{jMetalPyBenitez2019} framework for the implementation of the GA. \qusbt provides the encoding of the search individual, the definition of the fitness function that specifies the search problem, and the test assessment w.r.t. two types of failures for quantum programs~\cite{ourICST2021}; one test assessment relies on statistical tests to deal with the probabilistic nature of quantum programs.

We have validated \qusbt by testing ten faulty versions of five quantum programs. Overall, on average, \qusbt managed to almost always generate test suites in which at least 50\% of the test cases failed.

The rest of the paper is organized as follows. Sect.~\ref{sec:Background} presents background, Sect.~\ref{sec:preliminaries} explains preliminaries, and Sect.~\ref{sec:methodology} describes \qusbt's architecture and methodology. We present validation results in Sect.~\ref{sec:validation}, and conclude the paper in Sect.~\ref{sec:conclusion}.

\section{Background}\label{sec:Background}
A quantum bit (or a \textit{qubit} in short) is the basic unit of information in Quantum Computing (QC) in contrast to a \textit{bit} in classical computing. A qubit, like a bit, can take value 0 or 1. However, in \textit{superposition} (a special QC characteristic), a qubit can be in both 0 and 1 states at the same time. In contrast, a classical bit will be either in 0 or 1. A qubit in superposition has \textit{amplitudes ($\alpha$)} corresponding to its 0 and 1 states. An \textit{amplitude ($\alpha$)} is captured as a complex number consisting of a \emph{magnitude} and a \emph{phase}. The former shows the probability of a quantum program being in a specific state, whereas the latter represents the angle of the amplitude in the polar form, ranging from 0 to 2$\pi$ in radians. We can represent the state of a one-qubit program in the Dirac notation as:
\begin{equation*}
\small
\begin{array}{l}
\alpha_{0}\ket{0}+\alpha_{1}\ket{1}
\end{array}
\end{equation*}
$\alpha_{0}, \alpha_{1}$ are the \emph{amplitudes} associated to 0 and 1. The square of the absolute value of amplitude of a state indicates the probability of the program to be in that state. For example, $\ket{0}$'s magnitude is $|\alpha_{0}|^{2}$. The sum of all the magnitudes is 1, i.e., $\sum_{i=0}^1 |\alpha_{i}|^{2} = 1$.

\begin{figure}[!tb]
\centering
\includegraphics[width=1.0\columnwidth]{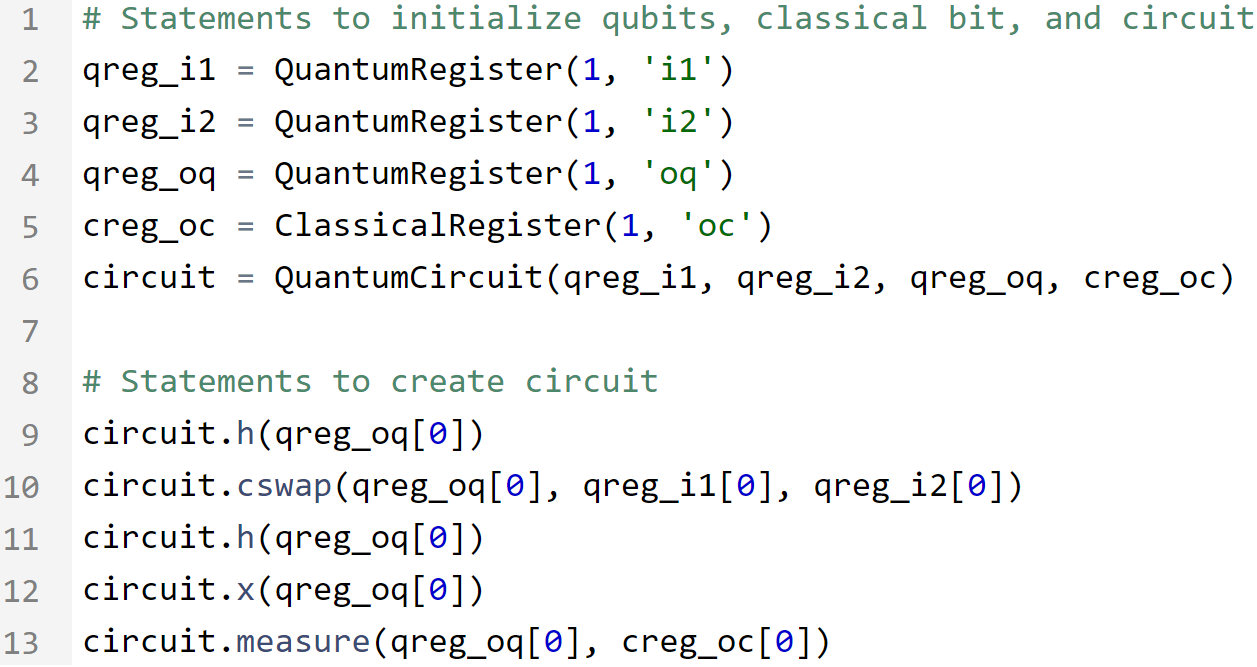}
\caption{Swap Test -- Qiskit Code}
\label{lst:code}
\end{figure}

\begin{figure}[!tb]
\centering
\includegraphics[width=0.9\columnwidth]{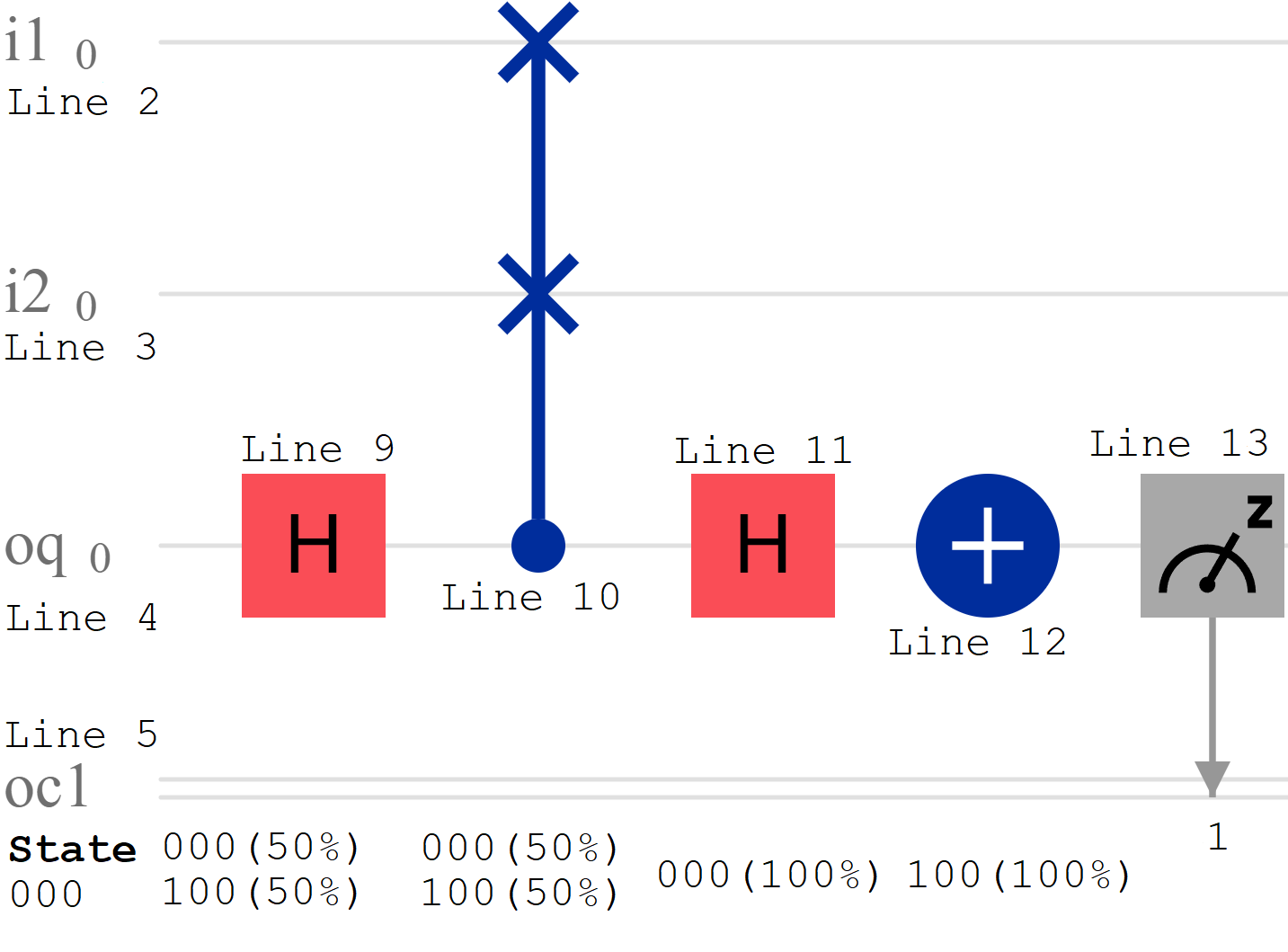}
\caption{Swap Test -- Circuit Diagram}
\label{fig:swapcircuit}
\end{figure}

In Fig.~\ref{lst:code}, we show an example of quantum program, called \textit{Swap Test}, encoded in Qiskit (taken from~\cite{QPBook}), alongside with its circuit in Fig.~\ref{fig:swapcircuit}.
The program compares two input qubits, i.e., \textit{i1} and \textit{i2} (initialized in Lines~2-3 in their respective quantum registers \textit{qreg\_i1} and \textit{qreg\_i2}). If \textit{i1} and \textit{i2} are equal, then output qubit (i.e., \textit{oq}, initialized in Line 4 in its register \textit{qreg\_oq}) will output the value 1 with 100\% probability when measured in the classical bit (i.e., \textit{oc}, initialized in its register \textit{creg\_oc} at Line~5). If \textit{i1} and \textit{i2} are different, then the probability decreases with the increased difference between the two inputs.

Lines~2-5 perform the necessary initialization of the two qubits and the classical bit, whereas Line~6 creates a quantum circuit for the initialized qubits and bit. By default, all the three qubits and the classical bit are initialized to 0, i.e., the program is in state 000.

Line~9 puts \textit{oq} in superposition with the Hadamard (\textit{H}) gate. As a result, the quantum program's state will be 000 and 100 with amplitudes of 0.707, i.e., 50\% of magnitude associated with each state. Note that the magnitude is calculated by taking the square of the absolute value of the amplitude. Since we put only \textit{oq} is superposition, it is the only qubit in both 0 and 1 states.

We apply the \textit{cswap} gate in Line~10 to swap the two inputs (i.e., the states of \textit{i1} and \textit{i2}) conditioned by the value of \textit{oq}. This means that if the value of \textit{oq} is 1, the states of \textit{i1} and \textit{i2} will be swapped. For the code shown in Fig.~\ref{fig:swapcircuit}, considering that \textit{oq} is initialized as 0, as a result, the program's state will remain the same after the swap. Line~11 applies the \textit{H} gate to \textit{oq} and, as a result, the program's state will be same as the initial: 000. We apply the not gate (i.e., the \textit{x} operation) on Line~12, which will transit the state of the program to 100. Finally, we measure \textit{oq} in the classical bit \textit{oc} in Line~13, which is 1, showing that \textit{i1} and \textit{i2} are equal.

\section{Preliminaries} \label{sec:preliminaries}

We here provide the minimal definitions regarding quantum programs and their testing.

\begin{definition}[Quantum program]\label{def:quantumProgram}
We identify with \quantumBitsSet the qubits of the quantum program \quantumProg. $\inputsSet \subseteq \quantumBitsSet$ defines the \emph{input} of the program, and $\outputsSet \subseteq \quantumBitsSet$ the \emph{output}. The input and output domains are defined as $\inputDom = \mathcal{B}^{|\inputsSet|}$ and $\outputDom = \mathcal{B}^{|\outputsSet|}$, respectively. A quantum program \quantumProg is then described by the function $\quantumProg \colon \inputDom \rightarrow 2^{\outputDom}$.
\end{definition}

Note that the definition shows that a quantum program can return different output values for the same input value.

We assume that the expected behavior of the program is specified by a program specification.

\begin{definition}[Program specification]\label{def:programSpec}
Given a quantum program $\quantumProg \colon \inputDom \rightarrow 2^{\outputDom}$, \progSpec is its \emph{program specification}. For a given input \testInputValue and possible output $h$, the program specification specifies the expected probability of occurrence of output $h$ for input \testInputValue (i.e., $\progSpecIO{i}{h} = p_h$).
\end{definition}




As for classical programs, testing a quantum program consists in executing it with some inputs. However, due to the non-deterministic nature of quantum programs, each input must be executed multiple times to observe the distribution of its outputs. Precise definitions are as follows.

\begin{definition}[Test input and test result]\label{def:test}
A \emph{test input} is defined as $\langle i, \numRepIndex\rangle$, where \testInputValue is an assignment to input qubits \inputsSet, and \numRepIndex is the number of times that \quantumProg must be run with \testInputValue. The \emph{test result} is then defined as $[\quantumProg(i),$ $\ldots,$ $\quantumProg(i)] =$ $[o_1,$ $\ldots,$ $o_{\numRepIndex}]$, where $o_j$ is the output returned by the program at the $j$th execution.
\end{definition}

The assessment of a test result requires to check the distribution of its output values. Given the result $[o_1,$ $\ldots,$ $o_{\numRepIndex}]$ returned for an input \testInputValue, these two types of failures can occur:
\begin{itemize}
\item \emph{Unexpected Output Failure} (\uof): it occurs when an output $o_j$ is not expected according to the program specification \progSpec, i.e., $\progSpecIO{\testInputValue}{o_j} = 0$;
\item \emph{Wrong Output Distribution Failure} (\wodf): it occurs when the returned output values follow a probability distribution significantly different from the one specified by the program specification. A \emph{goodness of fit} test with the Pearson's chi-square test~\cite{agresti2019introduction} is employed to check for the presence of this type of failure.
\end{itemize}

\section{QuSBT Tool Architecture and Methodology}\label{sec:methodology}

As for classical programs, also for quantum programs the generation of (possibly) failing tests is a relevant problem. To this aim, in~\cite{genTestsQPSSBSE2021}, we proposed a search-based approach called \qusbt, based on a genetic algorithm (GA). Given a quantum program \quantumProg, it generates a test suite consisting of \numTests test cases. \numTests is a parameter of the approach that can be set by the user.

In this paper, we describe how we have engineered that work in a tool that is usable by developers of quantum programs. We describe the input and configuration of \qusbt, followed by the process of test generation, execution, and assessment.

\subsection{Input and Configuration}\label{sec:inputAndConfiguration}

\begin{figure}[!tb]
\centering
\includegraphics[width=1\columnwidth]{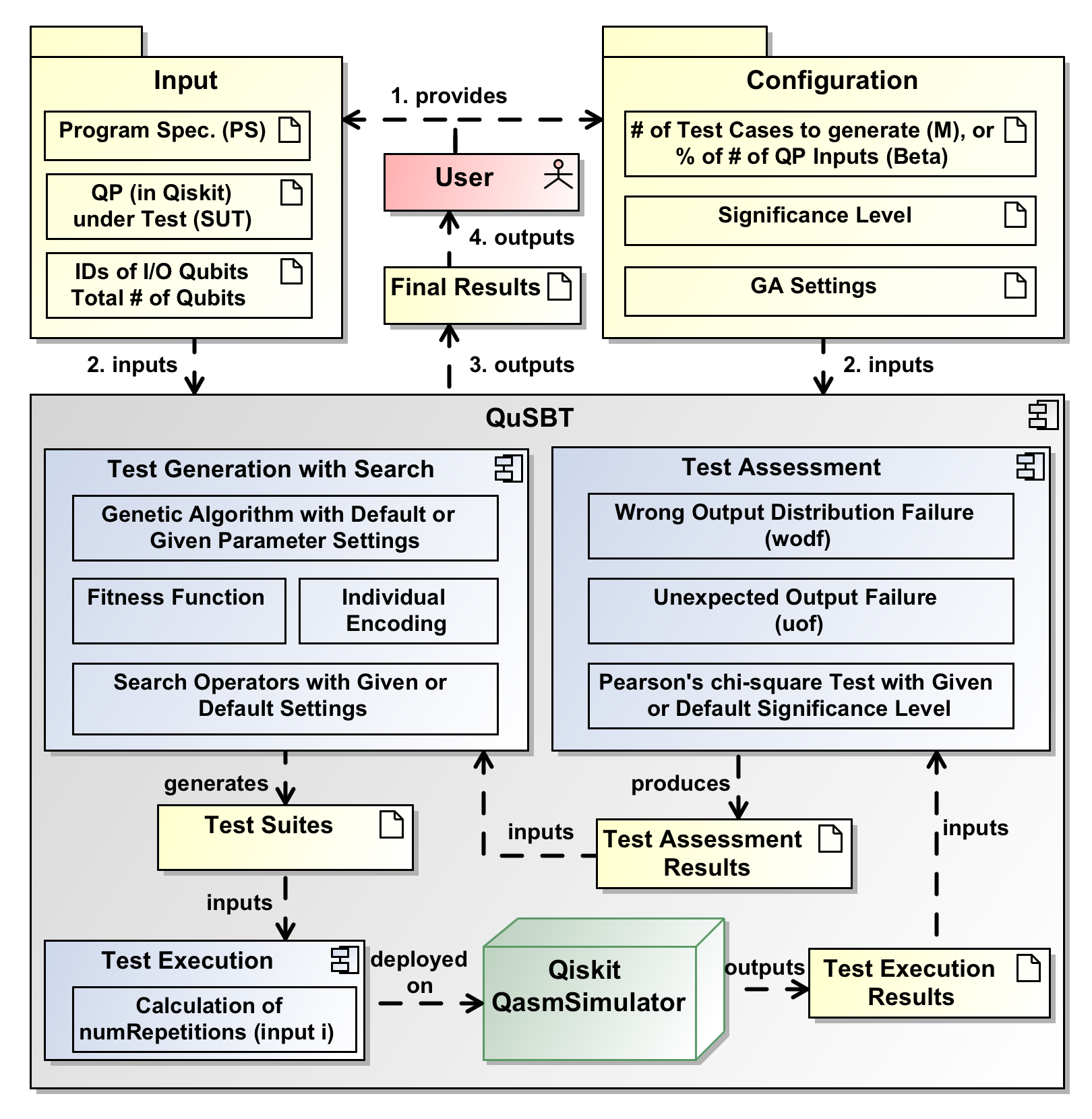}
\caption{Overview of the \qusbt Tool}
\label{fig:overview}
\end{figure}

To use \qusbt (see Fig.~\ref{fig:overview}), a user must provide \textit{Input Information}, which contains the quantum program under test (SUT), along with the list of the input and output qubits that it contains, the total number of qubits, and its program specification (PS) (see Def.~\ref{def:programSpec}).

Additionally, \qusbt requires users to select the number of tests \numTests to be added in each searched test suite. Users can specify \numTests, e.g., based on available budgets. However, selecting a value \numTests without considering the program under test might not be a good practice. So, \qusbt implements another way, i.e., selecting \numTests as the percentage \percInputs of the number of possible inputs \inputDom of the quantum program, i.e., $\numTests = \left \lceil{\percInputs \cdot |\inputDom|}\right \rceil$. In this way, users have the option to select the percentage \percInputs, rather than the absolute number \numTests.

Moreover, a user can also configure the GA or rely on the default GA settings in jMetalPy. These default settings are the binary tournament selection of parents, the integer SBX crossover (the crossover rate = 0.9), and the polynomial mutation operation being equal to the reciprocal of the number of variables. The population size is set as 10, and the termination condition is the maximum number of generations which is set as 50.
These default settings are provided in a template, starting from which a user can make their own {\it Input Information} file by changing the default settings, if needed. As discussed in Sect.~\ref{sec:preliminaries}, we employ the Pearson's chi-square test for checking failures of type \wodf. A user can specify the significance level to be used with the test; we set the default significance level at 0.01.


\subsection{Process of Test Generation, Execution, and Assessment}

We here describe how the search-based test generation is conducted in \qusbt.

An individual of the search is represented using integer variables $\overline{x} = [x_1,$ $\ldots,$ $x_{\numTests}]$; each $x_i$ represents a test case and it is defined over the input domain of the program \inputDom.

The goal of the search is to find an assignment $\overline{v} = [v_1,$ $\ldots,$ $v_{\numTests}]$ that maximizes the number of failing tests. Specifically, the fitness computation is as follows.
\begin{itemize}
\item For each assignment $v_j$ of the $j$th test:
\begin{itemize}
\item it computes the number of required repetitions (see Def.~\ref{def:test}) as $\numRepIndex_j = |\{h \in \outputDom \mid \progSpecIO{v_j}{h} \neq 0\}| \times 100$; the idea is that the number of required repetitions is proportional to the number of outputs that, according to the program specification, are expected to occur for input $v_j$;
\item it executes \quantumProg $\numRepIndex_j$ times with input $v_j$; the obtained {\it text execution results} are $[o_1,$ $\ldots,$ $o_{\numRepIndex_j}]$;
\item it assesses the correctness of the test execution result with respect to the two failure types (see Sect.~\ref{sec:preliminaries}):
\begin{itemize}
\item it checks if a failure of type \uof occurred, i.e., if an output that, according to the program specification \progSpec, should not occur has been produced; if this is the case, it sets flag $\fail_j$ to {\it true}, and the assessment for \wodf is not performed; otherwise,
\item it assesses \wodf by checking whether the frequency distribution of the measured output values (contained in the test execution results $[o_1,$ $\ldots,$ $o_{\numRepIndex_j}]$) follows the expected distribution specified in the program specification \progSpec. This is done by performing a \textit{goodness of fit test} with the Pearson's chi-square test, using the given or the default significance level. If the statistical test shows a significant difference, it sets flag $\fail_j$ to {\it true}, otherwise to {\it false}.
\end{itemize}
\end{itemize}
\item Given the assessments $\res = [\fail_1, \ldots, \fail_{\numTests}]$ of all the generated tests, the fitness function to be maximized is computed as follows:
\[\mathit{fitness}(\overline{v}) = |\{\fail_j \in \res \mid \fail_i = \mathit{true}\}|\]
\end{itemize}

The underlying GA runs till the specified number of generations. As output, the tool provides the test suite that maximizes the fitness function in two formats:
\begin{itemize}
\item as a list of input values, together with information of the returned output, and information on whether the test passed or failed;
\item as unit tests written in the {\tt unittest} framework\footnote{\url{https://docs.python.org/3/library/unittest.html}}; such format is useful for the user, who can re-run the tests, in particular the failing ones, while debugging the quantum program.
\end{itemize}

\section{Validation}\label{sec:validation}

For this paper, we conducted experiments to evaluate \qusbt with the following five quantum programs:
\begin{inparaenum}[(i)]
\item \bvName, the Bernstein-Vazirani cryptographic algorithm;
\item \qramName, an algorithm to access and manipulate quantum random access memory;
\item \iqftName, inverse quantum Fourier transform; 
\item \addSName, a mathematical operation in superposition;
\item \ceName, a conditional execution in superposition.
\end{inparaenum}
The total number of qubits of the programs ranges from 10 to 20, the number of gates from 15 to 60, and the circuit depth from 3 to 56. To assess whether \qusbt can find faults in the quantum programs, we created two faulty versions of each program, named as {\it \firstFaultyQP} and {\it \secondFaultyQP}.

Regarding the GA settings, we set the population size as 10, and we fixed 50 generations as the termination condition for the search. \qusbt requires to specify the number of tests $M$ to generate, and this can be done as percentage \percInputs of the size of the input domain (see Sect.~\ref{sec:inputAndConfiguration}); for these experiments, we used $\percInputs=5\%$.

Table~\ref{tab:results} summarizes the results for the faulty QPs in terms of the size of each test suite ({\it \numTestCases}), percentage of failing tests ({\it \percFailingTests}), simulation time ({\it \simulationTime}), i.e., the time to execute all the test cases with Qiskit QasmSimulator, and execution time of the search ({\it \executionTime}).
\begin{table}[!tb]
\centering
\caption{Experimental Results for Testing 10 Faulty QPs with \qusbt (\numTestCases: number of tests of the generated test suite; \percFailingTests: percentage of failing tests in the test suite; \simulationTime: simulation time; \executionTime: execution time of the search (excluding simulation))}
\label{tab:results}
\setlength{\tabcolsep}{3.5pt}
\begin{tabular}{lrrrr|rrrr}
\toprule
\multirow{2}{*}{\textbf{QP}} & 
\multicolumn{4}{c}{\textbf{\firstFaultyQP}} & 
\multicolumn{4}{c}{\textbf{\secondFaultyQP}} \\
\cline{2-9} & \numTestCases & \percFailingTests & \simulationTime (sec) & \executionTime (sec) & \numTestCases & \percFailingTests & \simulationTime (sec) & \executionTime (sec)\\
\midrule
\bvName & 52 & 75.0\% & 3201 & 7 &  52  & 76.9\% & 3151 & 7\\
\qramName &  26 & 57.7\% & 272 & 18 &  26 & 88.5\% & 238 & 5\\
\iqftName & 52 & 71.2\% & 9839 & 1504 &  52 & 82.7\% & 9837 & 1271\\
\addSName &  52 & 46.2\% & 732 & 28 &  52 & 50.0\% & 744 & 26\\
\ceName &  52 & 67.3\% & 968 & 22 &  52 & 73.1\% & 1003 & 20\\
\bottomrule
\end{tabular}
\end{table}
We can see that, while the search time is minimal, \qusbt spent most of the time executing test cases on the QPs using the simulator. Moreover, with the exception of \firstFaultyQP for \addSName, \qusbt managed to find at least 50\% of failing test cases in the generated test suite. It is not always possible to achieve a higher number of failing test cases, because, for instance, the total number of possible failing inputs of a quantum program could be much lower than \numTestCases. Nonetheless, we can conclude that \qusbt managed to find more than 50\% of failing test cases for most of the faulty QPs.

\section{Conclusion and Future Work}\label{sec:conclusion}
We presented a test generation tool (\qusbt) for quantum programs that uses a genetic algorithm to automatically find a test suite in which the number of failing test cases is maximized. \qusbt implements the encoding of the search individual, the fitness function describing the search problem, and the checking of passing and failing of tests w.r.t. two types of failures using a statistical test to deal with the probabilistic nature of quantum programs; moreover, it integrates with the IBM's Qiskit QasmSimulator. We presented the tool architecture and the detailed methodology, and conducted experiments with ten faulty versions of five quantum programs to validate the tool. In the future, we plan to integrate other search algorithms in \qusbt, and provide a methodology to guide users to configure and apply \qusbt.

\begin{acks}
This work is supported by the National Natural Science Foundation of China under Grant No. 61872182 and Qu-Test (Project\#299827) funded by Research Council of Norway. P. Arcaini is supported by ERATO HASUO Metamathematics for Systems Design Project (No. JPMJER1603), JST; Funding Reference number: 10.13039/501100009024 ERATO.
\end{acks}

\bibliographystyle{ACM-Reference-Format}
\bibliography{QuSBT_ICSE22_cameraReady}

\end{document}